\documentclass[a4paper,fleqn,usenatbib]{mnras}

\usepackage[T1]{fontenc}
\usepackage{ae,aecompl}

\usepackage{graphicx}	% Including figure files
\usepackage{amsmath}	% Advanced maths commands
\usepackage{amssymb}	% Extra maths symbols
\usepackage[usenames,dvipsnames]{xcolor}

\title[Gas squeezing during the merger of a SMBHB]{Gas squeezing during the merger of a supermassive black hole binary}
\author[A. Cerioli , G. Lodato and D. J. Price]{
Alice Cerioli$^{1}$\thanks{E-mail: alice.cerioli@gmail.com}, 
Giuseppe Lodato$^{1}$\thanks{E-mail: giuseppe.lodato@unimi.it} 
and Daniel J. Price$^{2}$
\\
$^{1}$Dipartimento di Fisica, Universit\`a degli Studi di Milano, Via Celoria 16, 20133, Milano, Italy\\
$^{2}$Monash Centre for Astrophysics and School of Physics \& Astronomy, Monash University, Vic 3800, Australia}

\date{Accepted 2016 January 4. Received 2015 December 21; in original form 2015 July 28}

\pubyear{2016}

\begin{document}
\label{firstpage}
\pagerange{\pageref{firstpage}--\pageref{lastpage}}
\maketitle

\begin{abstract}
We study accretion rates during the gravitational wave-driven merger of a binary supermassive black hole embedded in an accretion disc, formed by gas driven to the centre of the galaxy. We use 3D simulations performed with \textsc{phantom}, a Smoothed Particle Hydrodynamics code. Contrary to previous investigations, we show that there is evidence of a ``squeezing phenomenon'', caused by the compression of the inner disc gas when the secondary black hole spirals towards the primary. This causes an increase in the accretion rates that always exceed the Eddington rate. We have studied the main features of the phenomenon for a mass ratio $q = 10^{-3}$ between the black holes, including the effects of numerical resolution, the secondary accretion radius and the disc thickness.
With our disc model with a low aspect ratio, we show that the mass expelled from the orbit of the secondary is negligible ($< 5\%$ of the initial disc mass), different to the findings of previous 2D simulations with thicker discs.
The increase in the accretion rates in the last stages of the merger leads to an increase in luminosity, making it possible to detect an electromagnetic precursor of the gravitational wave signal emitted by the coalescence.
\end{abstract}


%%%%%%%%%%%%%%%%%%%%%%%%%%%%%%%%%%%%%%%%%%%%%%%%%%

%%%%%%%%%%%%%%%%% BODY OF PAPER %%%%%%%%%%%%%%%%%%

%%%%%%%%%%%%%%%%%%%%%%%%%%%%%%%%%%%%%%%%%%%%%%%%%%%%%%%%%%%%%%%%%%%%%%%%%%%%%%%%%%%%%%%%%%%%%%%%%%%
%%%%%%%%%%%%%%%%%%%%%%%%%%%%%%%%%%%%%%%%%%%%%%%%%%%%%%%%%%%%%%%%%%%%%%%%%%%%%%%%%%%%%%%%%%%%%%%%%%%
\section{Introduction}

There is clear evidence that most galaxies harbour a supermassive black hole \citep{1998AJ....115.2285M,2000ApJ...539L...9F}. The current hierarchical formation models state that our Universe evolves through the merger of smaller galaxies to form larger ones. During a galaxy-galaxy merger, the two supermassive black holes approach each other through dynamical friction and form a bound binary \citep{1980Natur.287..307B,2001ApJ...563...34M}.
Supermassive black hole pairs evolve by losing their angular momentum and energy by a combination of various processes, including interaction with a gas disc, until they reach a separation small enough that the main mechanism of shrinking can take over as gravitational radiation \citep{1964PhRv..136.1224P}.
Mergers of gas rich galaxies, through gravitational torques, can drive a large amount of gas in the core of the remnant galaxy, where the supermassive black hole pair resides. Gas rapidly settles into a thin accretion disc surrounding the two bound black holes. The gas rich environment may accelerate the orbital evolution of the binary \citep{2002ApJ...567L...9A,2005ApJ...630..152E,2007MNRAS.379..956D,2009MNRAS.393.1423C,2009MNRAS.398.1392L} and since the two black holes are expected to interact strongly with the surrounding material, we can investigate the possibility of identifying electromagnetic signatures of such interaction.

The theoretical modelling of electromagnetic signals from merging supermassive black hole binaries has developed in the last few years. Numerous mechanisms for eliciting an electromagnetic signature from coalescing binaries and recoiling remnants have been proposed \citep{2002ApJ...567L...9A,2005ApJ...622L..93M,2010PhRvD..81b4019S}; see \citet{2013CQGra..30x4007S} for a review.
This rapid theoretical development is justified by the hope of detecting gravitational waves signals by the improvement of ground based observatories and the prospects of space detectors, such as eLISA.
Ground based detections of galactic events will provide a test of the difficult task of extracting physical quantities from waveforms. 
However, the detection of gravitational waves from merging supermassive binaries will depend on the launch of space-based interferometers.

Independent of the detection of gravitational waves, merging supermassive black holes might be already detected in the electromagnetic domain if some peculiar feature indicating the merger can be identified.
In this paper we study the accretion rates of the black holes immediately before the final coalescence, to understand if the gas present between the two is squeezed and finally accreted very rapidly or if it manages to escape immediate accretion.
The possible increase of the accretion rates in the last stages of the merger can lead to an increase in luminosity, making it possible to detect an electromagnetic precursor of the gravitational waves signal emitted by the coalescence.
The coincident detection of both the gravitational wave and electromagnetic signatures allows for a possible high precision measurement of black hole masses, spins and the properties of the host environment.

Electromagnetic identification of the source would confirm the merger and accretion physics.
This idea was firstly investigated by \citet{2002ApJ...567L...9A}, using a one dimensional model. They found evidence of super-Eddington flares in the luminosity of the binary, caused by the rapid accretion of the inner disc.
Two dimensional simulations by \citet{2012MNRAS.423L..65B} suggested that the gas in the inner disc could actually flow across the secondary's gap through horseshoe orbits back to the outer disc, excluding the possibility of enhanced accretion luminosity.
Here we broaden the analysis providing a new picture of this phenomenon examining the case of thinner discs. We model the evolution of the inner accretion disc using three dimensional Smoothed Particle Hydrodynamics (SPH) simulations.

The paper is organized as follows: in Section \ref{sec:physprobl} we introduce the physics behind the gravitational decay of a supermassive black hole binary due to emission of gravitational radiation. In Section \ref{sec:overview} we give an overview on initial conditions and describe the system implemented in our numerical simulations. We show our results in Section \ref{sec:results}, where we illustrate the effects of increasing the resolution, changing the disc thickness and lowering the accretion radius of the smaller black hole. We discuss our results and draw conclusions in Sections \ref{sec:commanddisc} and \ref{sec:summandconcl}.

%%%%%%%%%%%%%%%%%%%%%%%%%%%%%%%%%%%%%%%%%%%%%%%%%%%%%%%%%%%%%%%%%%%%%%%%%%%%%%%%%%%%%%%%%%%%%%%%%%%
%%%%%%%%%%%%%%%%%%%%%%%%%%%%%%%%%%%%%%%%%%%%%%%%%%%%%%%%%%%%%%%%%%%%%%%%%%%%%%%%%%%%%%%%%%%%%%%%%%%
\section[]{Physical problem} \label{sec:physprobl}

When a binary is formed by unequal mass black holes, i.e. when the mass ratio $q=M_{\rm s}/M_{\rm p}$ between the secondary and primary mass is small, the secondary black hole exerts a tidal torque on the disc, pushing away gas from its orbit and creating a gap in the accretion disc around the primary. The disc is then split into an inner disc and an outer disc \citep{1979MNRAS.186..799L}.
This particular configuration is typical of protostellar discs, where the gap opening process is associated with planetary migration \citep{1986ApJ...309..846L}.
If the two black holes reach separations smaller than approximately $10^{-3}$ pc, they start to lose energy and angular momentum because of the emission of gravitational radiation \citep{1963PhRv..131..435P}. 
\citet{1964PhRv..136.1224P} found the equations for the average time variations of the eccentricity $e$, the separation $a$, the energy and the angular momentum of the binary in the generic case $e\neq 0$. 
If the binary reaches coalescence through the interaction with a disc, we expect any eccentricity to be damped by dissipative effects. We thus consider the case $e=0$, where
\begin{equation} \label{dadtcircgc}
\dot{a}(t)=-\frac{64}{5}\frac{G^3 M_{\rm p}^3q(1+q)}{c^5 a^3(t)},
\end{equation}
where $G$ is the gravitational constant and $c$ is the speed of light.
This differential equation can be solved analytically to give 
\begin{equation} \label{aaottau14}
a(t)=a_0\left(1-\frac{t}{\tau}\right)^{1/4},
\end{equation}
where $\tau$ is the time-scale to reach merger from an initial separation $a=a_0$, defined by
\begin{equation} \label{timescaletau}
\tau\equiv\frac{5}{256}\frac{c^5a_0^4}{G^3 M_{\rm p}^3q(1+q)}.
\end{equation}
Since the sign of (\ref{dadtcircgc}) is negative, the emission of gravitational waves causes the binary separation to decrease.
The time-scale in eq(\ref{timescaletau}) depends strongly on the separation $a_0$. For $q=10^{-3}$, $\tau$ becomes comparable to the Hubble time only at $a_0\approx 10^{-2} \rm pc$. At larger separations, other mechanisms need to be identified to allow the orbital decay of the binary. Dynamical friction against the stellar background is effective only for large separations $a\geq 1\rm pc$ \citep{1980Natur.287..307B,2001ApJ...563...34M}. The evolution from $a\approx 1\rm pc$ is probably due to the interaction with a gas disc \citep{2005ApJ...630..152E,2007MNRAS.379..956D,2009MNRAS.398.1392L,2015MNRAS.449.1118T}. Angular momentum exchange with a gas disc dominates the binary evolution down to the so-called ``decoupling radius'', where tidal torques equal the torque due to gravitational wave emission \citep{2005ApJ...622L..93M}.

Beyond decoupling the gas outside the secondary's orbit is unable to evolve on the rapid time-scale of gravitational decay so it remains frozen behind the secondary.
A simple estimate of the decoupling radius can be derived by equating the the gravitational decay migration rate to the viscous radial drift velocity, assumed to be the velocity of the secondary black hole in the disc dominated phase \citep{2002ApJ...567L...9A}, or by equating the gravitational decay time $\tau$ to the viscous time-scale $t_\nu$.
\citet{2005ApJ...622L..93M} self-consistently solve a stable standard $\alpha$-model for the binary-disc evolution, and equating the gravitational decay time-scale with the shrinking time of gas they estimate the decoupling radius as
\begin{equation} \label{eq:adecmp05}
\frac{a_{\rm dec}}{GM_{\rm p}/c^2}\approx 117 \left(\frac{\alpha}{0.1}\right)^{-0.34}\left(\frac{M_{\rm p}}{10^6 M_\odot}\right)^{0.08}\left(\frac{4q}{(1+q)^2}\right)^{0.42}Q,
\end{equation}
where $Q$ is a factor of order unity.
After the decoupling, $\tau < t_\nu$, so the inner edge of the outer disc cannot follow the secondary as it rapidly spirals in. 
Therefore, if between the black holes there is no gas, the final merger is expected to occur in vacuum \citep{2005ApJ...622L..93M}. 
On the other hand, if gas is still present at least around the primary black hole, the tidal torques of the secondary force the gas to be squeezed and rapidly accreted by the primary, causing a sudden enhancement of the accretion rate \citep{2002ApJ...567L...9A,2010MNRAS.407.2007C}. 
This would increase the disc's bolometric luminosity, constituting an electromagnetic precursor to the gravitational waves signal emitted by the final merger. This is the possibility considered by \citet{2002ApJ...567L...9A}, who performed one dimensional simulations to model the secondary migration in a gaseous disc. However, two dimensional simulations performed by \citet{2012MNRAS.423L..65B} suggest that during the gravitational decay the inner gas is progressively funnelled to the outer disc following horseshoe trajectories passing through the secondary orbit.
This would imply that there should be no significant increase in the binary luminosity prior to the merger with the choice of their initial conditions.
We discuss this issue again in Section \ref{sec:results}, in the light of the results of our simulations.

%%%%%%%%%%%%%%%%%%%%%%%%%%%%%%%%%%%%%%%%%%%%%%%%%%%%%%%%%%%%%%%%%%%%%%%%%%%%%%%%%%%%%%%
%%%%%%%%%%%%%%%%%%%%%%%%%%%%%%%%%%%%%%%%%%%%%%%%%%%%%%%%%%%%%%%%%%%%%%%%%%%%%%%%%%%%%%%
\section[]{Initial conditions} \label{sec:overview}

The evolution of the accretion disc and the black hole is modelled using Smoothed Particle Hydrodynamics (SPH), see for example \citet{1992ARA&A..30..543M,2005RPPh...68.1703M,2012JCoPh.231..759P} and \citet{2009NewAR..53...78R}.
SPH is a numerical method to solve the hydrodynamics equations in Lagrangian form, interpolating quantities over a finite set of $N$ point particles. 
Thus, resolution depends on the number of particles, and it is automatically higher in denser regions. 
The simulations presented in this paper have been carried out with the three dimensional code \textsc{phantom}, written by Daniel Price \citep{2010MNRAS.405.1212L,2010MNRAS.406.1659P}. 
\begin{table*}
 \centering
 \begin{tabular}{l | c | c | c | c | c | c | c | c | c }
 \hline
 SIMULATION & $\rm N_{\rm part}$ & $a_0$ & $\alpha_{\rm SS}$ & $H/R (R=R_{\rm in})$ & $R_{\rm in}=R_{\rm acc,p}$ & $R_{\rm acc,s}$ & $R_{\rm out}$ & $\alpha^{\rm AV}$ & $\langle h\rangle /H $\\
 \hline
 \emph{ref} & $5\cdot 10^5$ & 4.75 & 0.01 & 0.01 & 2 & 0.2 & 4.1 & 0.1114 & 0.8975  \\ % gwdisc_001 sim_5
 \emph{resol\_1} & $1\cdot 10^6$ & 4.75 & 0.01 & 0.01 & 2 & 0.2 & 4.1 & 0.1404 & 0.7124  \\ % gwdisc_007 sim_6
 \emph{resol\_2} & $2\cdot 10^6$ & 4.75 & 0.01 & 0.01 & 2 & 0.2 & 4.1 & 0.1768 & 0.5655 \\ % gwdisc_sep2 sim_7
 \emph{asp\_1} & $5\cdot 10^5$ & 4.75 & 0.01 & 0.015 & 2 & 0.2 & 4.1 & 0.1460 & 0.6850  \\ % gwdisc_006 sim_8
 \emph{asp\_2} & $5\cdot 10^5$ & 4.75 & 0.01 & 0.02 & 2 & 0.2 & 4.1 & 0.1768 & 0.5655  \\ % gwdisc_005 sim_9
 \emph{asp\_3} & $5\cdot 10^5$ & 4.75 & 0.01 & 0.05 & 2 & 0.2 & 4.1 & 0.3259 & 0.3068  \\ % gwdisc_004 sim_10
 \emph{asp\_4} & $5\cdot 10^5$ & 4.75 & 0.01 & 0.0095 & 2 & 0.2 & 4.1 & 0.1077 & 0.9287  \\ % gwdisc_011 sim_11
 \emph{sec\_1} & $5\cdot 10^5$ & 4.75 & 0.01 & 0.01 & 2 & 0.05 & 4.1 & 0.1114 & 0.8975  \\ % gwdisc_008 sim_12
 \emph{sec\_2} & $5\cdot 10^5$ & 4.75 & 0.01 & 0.01 & 2 & 0.01 & 4.1 & 0.1114 & 0.8975  \\ % gwdisc_009 sim_13
 \hline
 \end{tabular}
\caption{Summary of the simulations performed, showing the name of the simulation, the resolution given by the number of SPH particles $\rm N_{\rm part}$, the initial separation $a_0$ between the black holes, the desired Shakura-Sunyaev viscosity parameter $\alpha_{\rm SS}$, the value of the disc aspect ratio $H/R$ at the radius $R=R_{\rm in}$, the inner disc edge equal to the primary accretion radius $R_{\rm in}=R_{\rm acc,p}$, the secondary accretion radius $R_{\rm acc,s}$, the outer disc edge $R_{\rm out}$, the value of the artificial viscosity parameter $\alpha^{\rm AV}$ and the ratio $\langle h\rangle /H $, the approximated mean smoothing length over the disc scale height.}
\label{tab:simoverview}
\end{table*}

The system consists of an unequal mass black hole binary on a circular orbit (with masses $M_{\rm p}$ and $M_{\rm s}$ for the primary and secondary black hole, respectively) and of an initially axisymmetric circumprimary disc that lies in the same plane of the two black holes. 
The relative position $\textbf{r}$ of the secondary black hole with respect to the primary has components $r_x=a(t)\cos\vartheta(t)$ and $r_y=a(t)\sin\vartheta(t)$. The evolution of the angular position $\vartheta(t)$ is given by 
\begin{equation}
\dot{\vartheta}(t)=\sqrt{\frac{G(M_{\rm p}+M_{\rm s})}{a(t)^3}},
\end{equation}
which can be integrated to give
\begin{equation}
\vartheta(t)=-\frac{8\tau}{5}\sqrt{\frac{G(M_{\rm p}+M_{\rm s})}{a_0^3}}\left(1-\frac{t}{\tau}\right)^{5/8}.
\end{equation}
In the centre of mass system, the positions of the primary and the secondary black hole are, respectively, $\mathbf{r}_{\rm p}'=-M_{\rm s}/M_{\rm tot} \mathbf{r}$ and $\mathbf{r}_{\rm s}'=M_{\rm p}/M_{\rm tot}\mathbf{r}$.

%The mass ratio $q=M_{\rm s}/M_{\rm p}$ has been fixed to $q=10^{-3}$ for all the simulations. This choice is mostly dictated by the available computational resources, because a higher $q$ corresponds to a larger decoupling radius and a longer $\tau$ and consequently to a longer simulation. 
%The mass ratio is fixed during the computation; this means that the two black holes do not change their mass following accretion. 
Our simulations start with a circumprimary disc only. We decided to neglect the outer circumbinary disc because we always choose an initial separation that is smaller than the decoupling radius of the system. The outer disc would be frozen far behind the secondary, so it would not affect the dynamics inside the secondary's orbit.
However, during the simulations, in some cases we can see that a small amount of gas (the exact fraction will be discussed later) is funnelled outside the orbit of the secondary, forming a low mass outer disc.

The mass ratio $q=M_{\rm s}/M_{\rm p}$ has been fixed to $q=10^{-3}$ for all of the simulations. This choice is mostly dictated by the available computational resources. A higher $q$ corresponds to a larger decoupling radius, but this is roughly balanced by a corresponding decrease in the number of orbits required to merge from a given initial separation. However, a higher $q$ also leads to a much stronger perturbation to the circumprimary disc, since a higher mass secondary produces a wider gap in the disc. This means that higher mass secondaries must be placed further away in order to avoid a strong perturbation to the initial conditions, leading to a considerably longer computational time to merger. 
The mass ratio is fixed during the computation; this means that the two black holes do not change their mass following accretion.

The accretion radius of the primary black hole $R_{\rm acc,p}$ is taken to be equal to the inner radius of the accretion disc $R_{\rm in}$, so that all mass approaching the black hole will be eventually accreted.  
In our simulations we chose $R_{\rm in}=2GM_{\rm p}/c^2$. 
Then, we can assume that this is the position of the last stable orbit for a non-maximally rotating black hole. However, no effects related to a Kerr black hole were taken into account in the implementation. The secondary accretion radius $R_{\rm acc,s}$ is chosen for computational efficiency, and we will discuss the effect of varying it in Section \ref{sec:varysecaccrrad}. In \textsc{phantom}, when particles approach a black hole by a distance less than its accretion radius, they are considered accreted. 
The mass and the momentum of the particles are not added to the black holes because we are treating the binary as an external potential. However, this is negligible anyway since the black holes are much more massive than the gas disc we considered.

The gas follows a locally isothermal equation of state, where the sound speed $c_{\rm s}$ is described by the power law $c_{\rm s}=c_{\rm s,0}R^{-\gamma}$ with $\gamma=0.75$ and where $R$ is the radial distance from the centre of mass of the binary in cylindrical coordinates. The disc surface density is modelled as $\Sigma=\Sigma_0 R^{-\delta}$, with $\delta=1.5$.

We model a \citet{1973A&A....24..337S} viscosity by using the SPH artificial viscosity formalism. The value of the viscosity parameter is $\alpha_{\rm SS}=0.01$, set by the relation
\begin{equation} \label{eq:artificialviscosity}
\alpha_{\rm SS}=\frac{1}{10}\alpha^{\rm AV} \frac{\langle h \rangle}{H},
\end{equation}
where $\alpha^{\rm AV}$ is the coefficient of artificial viscosity \citep{2010MNRAS.405.1212L}. The other two quantities in equation (\ref{eq:artificialviscosity}) are $\langle h \rangle$, the azimuthally averaged smoothing length of SPH particles and $H=c_{\rm s}/\Omega$, the disc thickness defined by the sound speed and the Keplerian velocity $\Omega$.
The chosen disc model with $\delta=1.5$ and $\gamma=0.75$ ensures that the disc is uniformly resolved since we have:
\begin{equation}
h\propto \rho ^{-1/3} \propto \left(\frac{\Sigma}{H}\right)^{-1/3} \propto R^{3/4}.
\end{equation}

\noindent The smoothing length $h$ is thus proportional to the disc thickness $H\propto R^{3/4}$, so that the ratio $\langle h \rangle /H$ is constant with radius.
The ratio $\langle h\rangle /H$ is a good estimator of how well our simulation resolves vertical disc scale height. 

For our chosen value of $\alpha$ and $q$, the decoupling radius would be $a_{\rm dec}\approx 25GM_{\rm p}/c^2$, with little dependence on the mass scale.
We have experimented with several choices of the initial separation $a_0$. We have noticed that during the early evolution of the system there is no or little evolution in the accretion rate onto the primary until the binary separation decreases to $\approx 5GM_{\rm p}/c^2$. In order to save computational time, we decided to start our simulations at $a_0=4.75 GM_{\rm p}/c^2$. We have also performed a number of simulations with a fixed, non decaying binary, in order to assess the radius at which the circumprimary disc is truncated by the secondary's tidal force.
For an initial separation $a_0=4.75 GM_{\rm p}/c^2$, this turns out to be $R_{\rm out}= 0.86 a_0$, that we take as outer disc radius in all our simulations.

In the following, we use code units for mass, length and time given by:
\begin{align} \label{eqs:codeunitschap5}
M_0&=M_{\rm p}, \nonumber \\
R_0&=\frac{GM_{\rm p}}{c^2}, \\
T_0&=\frac{GM_{\rm p}}{c^3}. \nonumber
\end{align}

Due to finite size of black hole accretion radii, the final merger will occur at a time 
\begin{equation}
\tau_{\rm m}=\tau (a_0)-\tau (\bar{a})
\end{equation}
where $\bar{a}=R_{\rm acc,p}+R_{\rm acc,s}$.
We refer to this as the merger time in the following discussion.
For example, considering an initial separation $a_0=4.75$, $R_{\rm acc,p}=2$ and $R_{\rm acc,s}=0.2$, the decay time is $\tau_{\rm m}=9475.7$, corresponding to $\approx 54$ days and to $145.7$ $T_{\rm orb}(a_0)$, where $T_{\rm orb}(a_0)$ is the orbital period of the secondary black hole at the initial separation.

Table \ref{tab:simoverview} summarizes the initial conditions, giving the name of the simulation, the resolution (given by the number of SPH particles $N_{\rm part}$), the initial separation $a_0$ between the black holes, the desired Shakura-Sunyaev viscosity parameter $\alpha_{\rm SS}$, the value of the disc aspect ratio $H/R$ at $R_{\rm in}$, the inner edge of the disc (equal to the primary accretion radius $R_{\rm in}=R_{\rm acc,p}$), the secondary accretion radius $R_{\rm acc,s}$, the disc outer edge $R_{\rm out}$, the value of the artificial viscosity parameter $\alpha^{\rm AV}$ set by the code and the ratio $\langle h\rangle /H $, the average smoothing length over the disc scale height.

%%%%%%%%%%%%%%%%%%%%%%%%%%%%%%%%%%%%%%%%%%%%%%%%%%%%%%%%%%%%%%%%%%%%%%%%%%%%%%%%%%%%%%%%
%%%%%%%%%%%%%%%%%%%%%%%%%%%%%%%%%%%%%%%%%%%%%%%%%%%%%%%%%%%%%%%%%%%%%%%%%%%%%%%%%%%%%%%%
\section{Results} \label{sec:results}
\subsection{Reference run}

We first present the results of our reference calculation, where we set $H/R(R=R_{\rm in})=0.01$ (Table \ref{tab:simoverview}). Then, we study the effects of increasing the resolution (Section \ref{sec:varresolution}), changing the disc aspect ratio (Section \ref{sec:vardiscthick}) and lowering the secondary accretion radius (Section \ref{sec:varysecaccrrad}).

The reference simulation \emph{ref} will always be considered as the starting point for the other simulations.
However, since simulation \emph{ref} has been run with low resolution ($N_{\rm part}=5\cdot 10^5$), we also present the results of the simulation \emph{resol\_2} with $N_{\rm part}=2\cdot 10^6$.
The qualitative results are the same in both simulations.

%%%%%%%%%%%%%%%%%%%%%%%%%%%%%%%%%%%%%%%%%%%%%%%%%%%%%%%%%%%%%%%%%%%%%%%%%%%%%%%%%%%%%%%%%%%%%%%%%%%
%%%%%%%%%%%%%%%%%%%%%%%%%%%%%%%%%%%%%%%%%%%%%%%%%%%%%%%%%%%%%%%%%%%%%%%%%%%%%%%%%%%%%%%%%%%%%%%%%%%
\subsubsection{The squeezing phenomenon}

\begin{figure}
	\centering
        \includegraphics[width=\columnwidth]{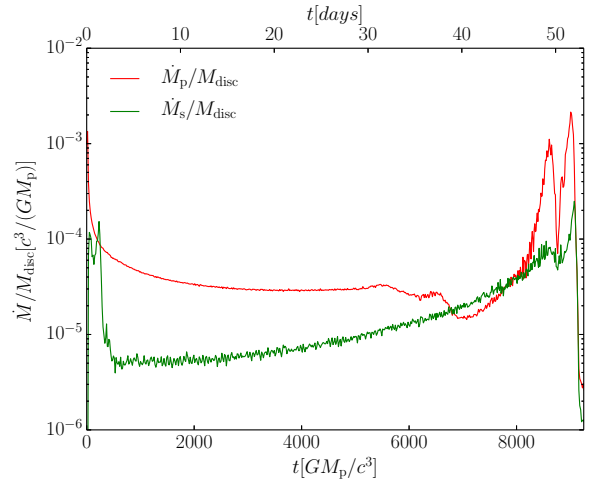}
	\caption[Accretion rates of simulation \emph{resol\_2}]{Primary and secondary accretion rates of simulation \emph{resol\_2}. Time $t$ is expressed in code units and the accretion rates are normalised with disc mass $M_{\rm disc}$. The sharp spikes mark episodes of strong and sudden accretion. The upper axis shows time expressed in days for a primary mass $M_{\rm p}=10^8 M_\odot$.}
	\label{fig:dmdt12resol2ref}
\end{figure}

In our reference simulations the secondary black hole rapidly shrinks towards the primary, squeezing the inner disc gas. 
Figure \ref{fig:dmdt12resol2ref} shows the primary and secondary accretion rates $\dot{M}_{\rm p}$ and $\dot{M}_{\rm s}$, normalised by the initial disc mass $M_{\rm disc}$. After a short initial transient where the initially axisymmetric disc readjusts to the imposed tidal torques, the accretion rates approach a quasi steady configuration where $\dot{M}_{\rm p}$ is almost one order of magnitude higher than $\dot{M}_{\rm s}$. As the binary shrinks, the secondary accretion rate starts to grow approaching the primary's value. As the merger eventually occurs, there is a strong spike in the primary rate $\dot{M}_{\rm p}$, that reaches a maximum value almost two orders of magnitude above the quiescent value.
The inner disc shows a spiral structure due to the tidal interaction with the secondary. This implies that the subsequent accretion of regions of different density modulates the accretion rates and the surface density curves.
This explains the presence of two maxima in the plot of the accretion rate of the primary black hole $\dot{M}_{\rm p}$ (Figure \ref{fig:dmdt12resol2ref}). 
\begin{figure*}
 \centering
 \includegraphics[width=\textwidth]{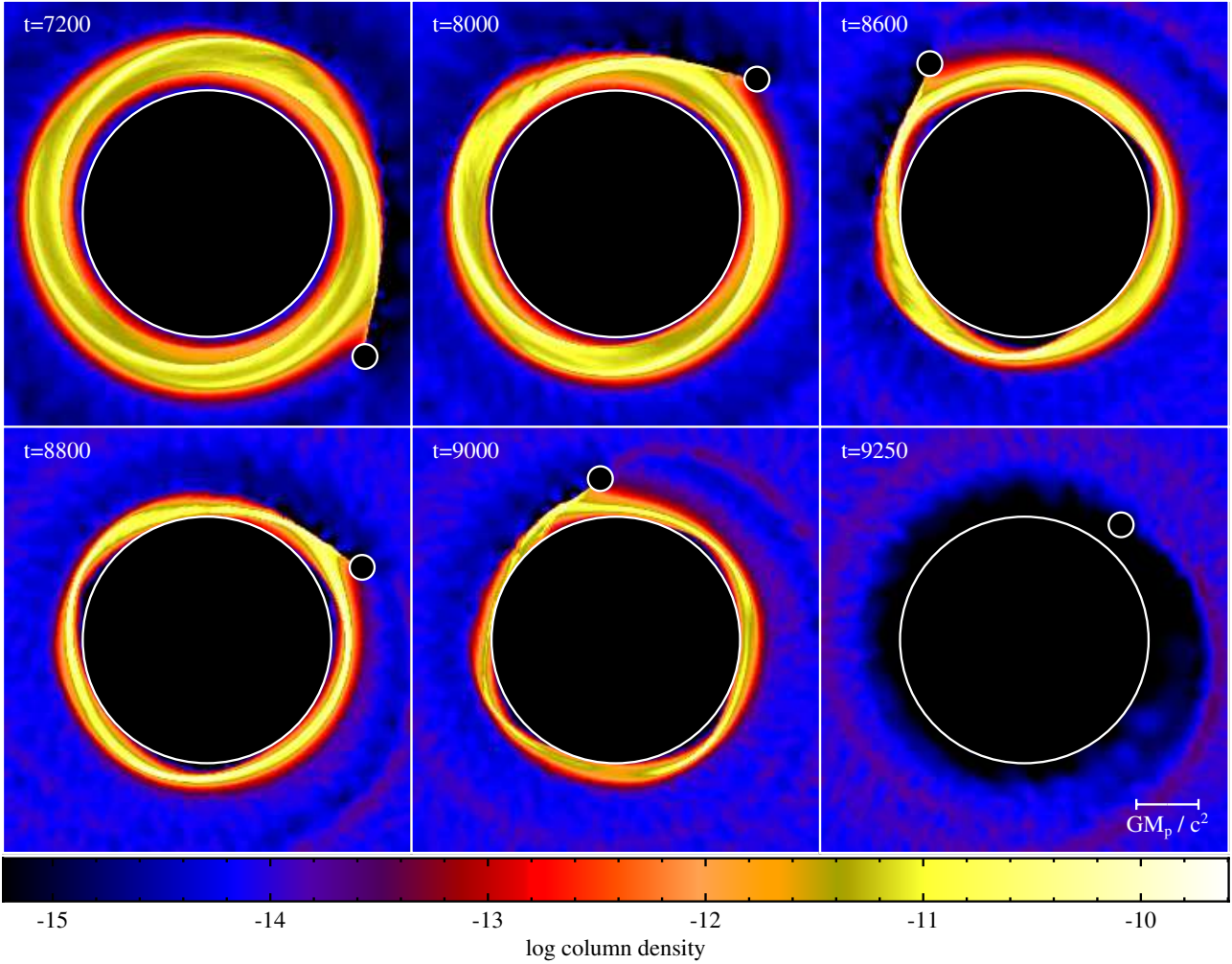}
\caption[\emph{resol\_2}: Snapshot of simulation]{Snapshots of column density in simulation \emph{resol\_2}. The two black holes are indicated by circles, whose size is proportional to the accretion radius ($R_{\rm acc,p}=2GM_{\rm p}/c^2$ and $R_{\rm acc,s}=0.2GM_{\rm p}/c^2$). In the snapshots time is expressed in code units. The binary separation decreases from $a=3.44GM_{\rm p}/c^2$ at $t=7200GM_{\rm p}/c^3$ to $a=2.43GM_{\rm p}/c^2$ at $t=9250GM_{\rm p}/c^3$ }
\label{sim7_snapshotmovie}
\end{figure*}
\begin{figure}
	\centering
        \includegraphics[width=\columnwidth]{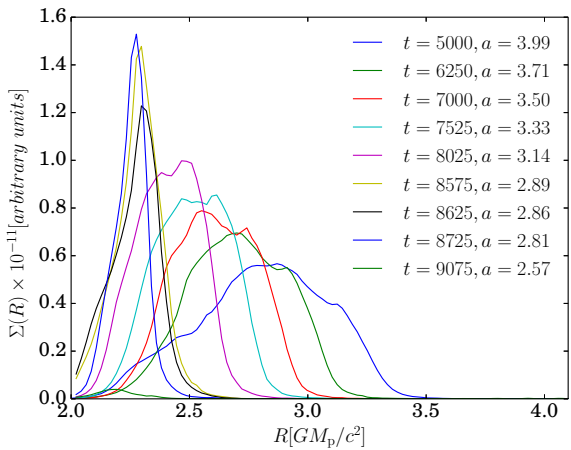}
	\caption[\emph{resol\_2}: Snapshot of surface density]{Snapshots of the surface density $\Sigma(R)$ for the simulation \emph{resol\_2}. For each snapshot, the time $t$ and the separation $a$ (that roughly coincides with the position $R_{\rm s}$ of the secondary) are indicated. Gas is squeezed by the rapid decay of the secondary black hole towards the primary. }
	\label{sim7_snapshotdensity}
\end{figure}

During the decay, the secondary black hole sweeps up the gas towards the primary, causing the formation of spikes in the disc surface density.
Figure \ref{sim7_snapshotdensity} shows the evolution of the azimuthally averaged surface density $\Sigma(R)$ at different times.
The surface density increases by a factor of 3 as the binary separation decreases from about $4$ to $2.6$.
Figure \ref{sim7_snapshotmovie} shows snapshots of column density of the simulation in the plane $xy$. The colour scale has been chosen in order to see the low mass outer disc that formed outside secondary's orbit. 
The snapshots at times $t=8600$ and $t=9000$ coincide with the two spikes in the accretion rate $\dot{M}_{\rm p}$.

The disc surface density spikes are caused by the rapid shrinking of the secondary. The inner disc is unable to viscously respond to the increasingly rapid infall of the black hole, so the gas is squeezed and pushed inwards. 
The mechanism is the same as for planets opening gaps. The interaction of the secondary with the inner disc results in an exchange of angular momentum between the satellite and the gas. Gas particles lose angular momentum and shift onto inner orbits.

The forced accretion of the gas on to the primary black hole, marked by peaks in the accretion rate, causes an increase of the disc accretion luminosity just prior to the merger. 

\begin{figure}
	\centering
	\includegraphics[width=\columnwidth]{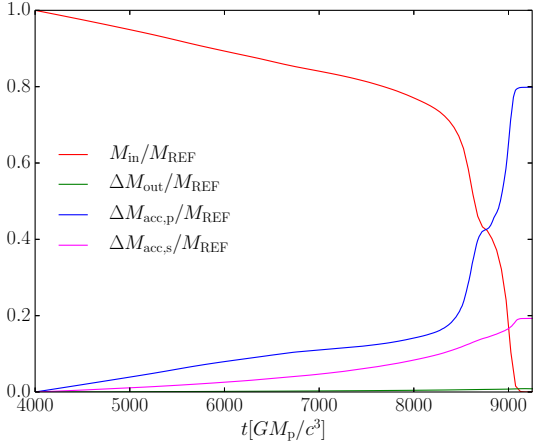}
	\caption[\emph{resol\_2}: Inner disc mass evolution]{Simulation \emph{resol\_2}: Inner disc mass evolution in the last part of the simulation. $M_{\rm in}$ represents the inner disc mass. $\Delta M_{\rm acc,p}$ and $\Delta M_{\rm acc,s}$ are the mass accreted by the primary and by the secondary in the time interval considered. $\Delta M_{\rm out}$ is the increase in the outer disc mass starting from time $t=4000$. $M_{\rm REF}$ is the inner disc mass at time $t=4000$. From the plot we can infer that only the 0.8\% of the disc mass is expelled outside (green curve) and that the inner disc mass is accreted almost completely by the two black holes. }
	\label{fig:sim7inn}
\end{figure}

Figure \ref{fig:sim7inn} shows the mass accreted by the primary, $\Delta M_{\rm acc,p}/M_{\rm REF}$ (blue curve), where $M_{\rm REF}$ is the inner disc mass at time $t=4000$ (no significant evolution of the accretion rates is observed before this time, see Fig. \ref{fig:dmdt12resol2ref}).
This shows the behaviour inferred from $\dot{M}_{\rm p}$, growing suddenly in two steps, corresponding to the two spikes in $\dot{M}_{\rm p}$.
At the end of the calculation, $79.9\%$ of the inner disc mass $M_{\rm REF}$ is accreted by the primary black hole, while $19.3\%$ is accreted by the secondary.
Only $0.8\%$ of the disc mass is expelled outside the binary orbit. 
This is a remarkable result since this means that the merger of two black hole happens in a relatively gaseous environment, in contrast to what was found by \citet{2012MNRAS.423L..65B} in their thicker disc case. 

%%%%%%%%%%%%%%%%%%%%%%%%%%%%%%%%%%%%%%%%%%%%%%%%%%%%%%%%%%%%%%%%%%%%%%%%%%%%%%%%%%%%%%%%%%%%%%%%%%%
%%%%%%%%%%%%%%%%%%%%%%%%%%%%%%%%%%%%%%%%%%%%%%%%%%%%%%%%%%%%%%%%%%%%%%%%%%%%%%%%%%%%%%%%%%%%%%%%%%%
\subsubsection{Horseshoe orbits}

\begin{figure}
	\includegraphics[width=\columnwidth]{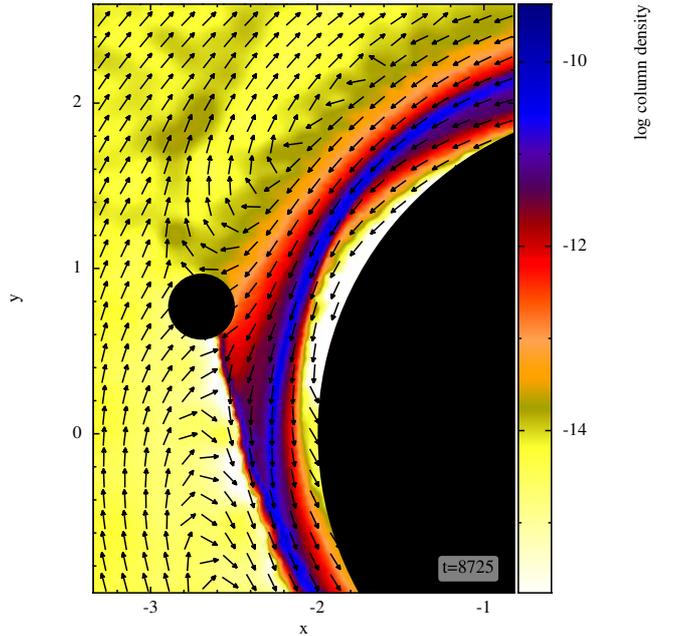}
	\caption[Horseshoe orbits, \emph{resol\_2}]{Horseshoe orbits, simulation \emph{resol\_2}. Arrows indicate particles velocity in the reference system co-rotating with the secondary black hole. Ahead and behind the secondary black hole we can see the inversion points, where the direction of particles motion changes. Particles located inside the orbit of the secondary's orbit are moving quicker than the black hole, while the particles outside are slower. Particles located in the corotation region exert horseshoe orbits and they can end up in a different region of the disc after the execution of a turn if the horseshoe region is asymmetric. }
	\label{fig:sim7_horseshoe}
\end{figure}

Figure \ref{fig:sim7_horseshoe} shows a close-up of the region near the secondary black hole in Cartesian coordinates at time $t=8725$, with velocities shown in the frame rotating with the secondary black hole. Hence arrows indicate the direction of particle motion with respect to the secondary. 
Mass is funnelled outside the orbit of the secondary black hole through horseshoe orbits \citep{1999ssd..book.....M}, as seen in the reference frame of the rotating smaller black hole.
Horseshoe orbits are equilibrium orbits that lie around the secondary orbit and encircle the Lagrangian points $L_3$, $L_4$ and $L_5$ in the equipotential surfaces of the effective potential of a rotating binary system. 

In the inertial frame the secondary and the disc rotate counterclockwise around the primary, located near the centre of mass, since we choose a small mass ratio.
Particles located in the inner disc, in Keplerian rotation, are moving more rapidly with respect to the secondary, so, in the velocity map, the arrows show a counterclockwise motion. By contrast, the slower particles in the outer disc are receding from the secondary.
The transfer of gas is possible because the horseshoe orbits are not closed as predicted by theory, but instead the orbits are asymmetric due to the inward drift of the secondary, meaning that the width of the horseshoe region is larger behind the black hole, similar to what occurs for planets in discs \citep{1999ApJ...526.1001L}.
The streamlines near the inversion point behind the black hole do not bring back gas in the horseshoe region, but channel it outside. 
The inverse process, i.e. material transferred from the outer disc to the inner is also possible. However, only gas located near the disc edge can embark on horseshoe orbits to be funnelled through the gap, so the process is unidirectional due to the very low mass of the outer disc and to the presence of the gravitational decay. 
The outer disc would also be low mass in a realistic scenario, since the outer circumbinary disc would be frozen far away from the black holes in the post-decoupling phase. 
The mass located in the outer disc is thus only the mass transferred outwards from the inner disc.

%%%%%%%%%%%%%%%%%%%%%%%%%%%%%%%%%%%%%%%%%%%%%%%%%%%%%%%%%%%%%%%%%%%%%%%%%%%%%%%%%%%%%%%%%%%%%%%%%%%
%%%%%%%%%%%%%%%%%%%%%%%%%%%%%%%%%%%%%%%%%%%%%%%%%%%%%%%%%%%%%%%%%%%%%%%%%%%%%%%%%%%%%%%%%%%%%%%%%%%
\subsection{Effect of resolution} \label{sec:varresolution}

\begin{figure}
 \centering
  \includegraphics[width=\columnwidth]{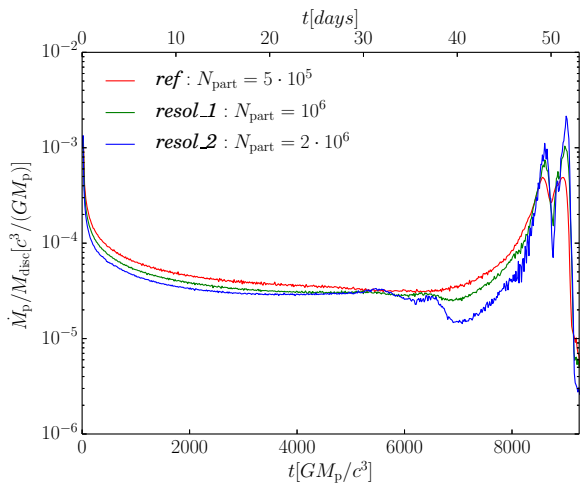}
\caption[Accretion rates at different resolutions]{ Primary accretion rates at different resolutions. The simulation \emph{resol\_2} presents the higher accretion rate. The blue curve reach a spikes that is a factor $72$ higher than the mean value for time inferior to $t=5000$. The spike of the primary accretion rate of \emph{resol\_2} is 4.5 times the spike of the resolution \emph{ref}, the one with lower resolution. The upper axis shows time expressed in days for a primary mass $M_{\rm p}=10^8 M_\odot$. 
$\dot{M}_{\rm p}$ is normalised by the initial disc mass $M_{\rm disc}$.
}
\label{fig:sim567res_accr}
\end{figure}

Figure \ref{fig:sim567res_accr} shows the accretion rates of the primary mass for three different resolutions (see Table \ref{tab:simoverview}).
The two spikes in $\dot{M}_{\rm p}$ are present at each resolution, indicating this is not a numerical artefact.

The simulation \emph{resol\_2} ($N_{\rm part}=2\cdot 10^6$) has the most significant enhancement in the primary accretion rate, shown in Figure \ref{fig:sim567res_accr} in units of the initial disc mass $M_{\rm disc}1$.
$\dot{M}_{\rm p}/M_{\rm disc}$ grows from $3\cdot 10^{-5}$ to $1.1\cdot 10^{-3}$ (by a factor of $37$) in the first spike and reaches $2.15 \cdot 10^{-3}$ (a factor of $72$ higher) in the second spike.
Similarly, the secondary accretion rate spike is $50$ times higher than the initial value (Figure \ref{fig:dmdt12resol2ref}). 

The huge enhancements in the accretion rates can be considered the signature of a black hole merger in a gaseous environment and later we comment on the consequences this could have on the accretion luminosity of the source in a realistic scenario (Section \ref{sec:physicalunits}).
As already observed, the squeezing mechanism can lead to a possible observable electromagnetic signal preceding the gravitational waves that will be intercepted by next generation detectors. 

%%%%%%%%%%%%%%%%%%%%%%%%%%%%%%%%%%%%%%%%%%%%%%%%%%%%%%%%%%%%%%%%%%%%%%%%%%%%%%%%%%%%%%%%%%%%%%%%%%%
%%%%%%%%%%%%%%%%%%%%%%%%%%%%%%%%%%%%%%%%%%%%%%%%%%%%%%%%%%%%%%%%%%%%%%%%%%%%%%%%%%%%%%%%%%%%%%%%%%%
\subsection{Effect of disc thickness} \label{sec:vardiscthick}

\begin{figure}
 \centering
  \includegraphics[width=\columnwidth]{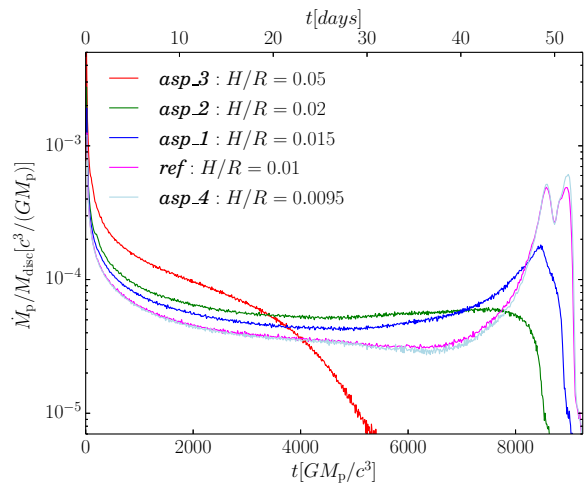}
\caption[Accretion rates with different disc thickness.]{ Accretion rates $\dot{M}_{\rm p}/M_{\rm disc}$ for different values of $H/R$. In the simulations \emph{asp\_2} and \emph{asp\_3} all the inner disc mass is accreted or expelled before the establishment of the squeezing mechanism, that on the contrary is visible in the other cases. Upper axis shows time in days for a primary mass $M_{\rm p}=10^8 M_\odot$.}
\label{fig:simhr_accr}
\end{figure}
\begin{table}
 \centering
 \begin{tabular}{l | c | c | c | c | c}
 \hline
 SIMULATION  & \emph{asp\_4}  & \emph{ref} &\emph{asp\_1} &  \emph{asp\_2} & \emph{asp\_3}\\
 \hline
 $H/R (R=R_{\rm in})$ & $0.0095$ &  $0.01$ & $0.015$ & $0.02$ & $0.05$ \\[1mm]
 $M_{\rm acc(TOT)}/M_{\rm disc}$ & $0.980$ & $0.978$ & $0.966$ & $0.954$ & $0.949$\\[1mm]
 $M_{\rm out}/M_{\rm disc}$ & $0.020$ & $0.022$ & $0.034$ & $0.046$ & $0.051$ \\[1mm]
 $\tilde{t}$ & $9250$ & $9250$ & $9250$ & $8875$ & $7500$\\
 \hline
 \end{tabular}
 \caption[Disc masses for different values of disc thickness]{Total accreted mass $M_{\rm acc(TOT)}$ and mass funnelled outside $M_{\rm out}$ for different values of $H/R$. Masses are expressed as fraction of the total initial disc mass. For each simulation we define the time $\tilde{t}$ at which the inner disc mass $M_{\rm in}/M_{\rm disc}$ decreases below $10^{-4}$. }
	\label{tab:hrcomp_maccmout}
\end{table}
The availability of a three dimensional code such as \textsc{phantom} allows us to study the dependence of the squeezing phenomenon on the disc thickness.
The thickness of the disc may affect the quantity of gas expelled outside the orbit of the secondary.

We first try to increase the disc thickness (disc aspect ratio) in simulations \emph{asp\_1}, \emph{asp\_2}, \emph{asp\_3} and to decrease it in simulation \emph{asp\_4}. Their initial conditions are listed in Table \ref{tab:simoverview}; they differ only for the chosen value of $H/R$.
 
Figure \ref{fig:simhr_accr} shows the accretion rates of the primary black hole for different values of the disc aspect ratio $H/R$, normalised by initial disc mass $M_{\rm disc}$.
In simulations \emph{asp\_2} and \emph{asp\_3}, corresponding to the values of $H/R=0.02$ and $H/R=0.05$, respectively, we do not see any effects of the forced compression because all mass is swallowed before it could be squeezed by the secondary, due to the increased viscosity. 

The accretion rates $\dot{M}_{\rm p}$ and $\dot{M}_{\rm s}$ increase in magnitude with increasing disc thickness in the first part of the simulations. For the thinner discs of \emph{ref} and \emph{asp\_4}, $\dot{M}_{\rm p}$ presents the two spike feature encountered before, while the gas in \emph{asp\_1} is too little to produce a notable peak.

In Table \ref{tab:hrcomp_maccmout} we show the total accreted mass $M_{\rm acc(TOT)}$ and the mass funnelled outside $M_{\rm out}$ for different values of $H/R$. Masses are  expressed as fraction of the total initial disc mass $M_{\rm disc}$. For each simulation we define the time $\tilde{t}$ at which the inner disc mass $M_{\rm in}/M_{\rm disc}$ decreases below $10^{-4}$. From Table \ref{tab:hrcomp_maccmout} we can deduce the following trend: when $H/R$ decreases, $M_{\rm acc(TOT)}$ increases and consequently, $M_{\rm out}$ decreases. In other words, if the disc is thick, gas can more easily flow outwards through the gap. 

%%%%%%%%%%%%%%%%%%%%%%%%%%%%%%%%%%%%%%%%%%%%%%%%%%%%%%%%%%%%%%%%%%%%%%%%%%%%%%%%%%%%%%%%%%%%%%%%%%%
%%%%%%%%%%%%%%%%%%%%%%%%%%%%%%%%%%%%%%%%%%%%%%%%%%%%%%%%%%%%%%%%%%%%%%%%%%%%%%%%%%%%%%%%%%%%%%%%%%%
\subsection{Effect of the accretion radius} \label{sec:varysecaccrrad}

Here we present the effects of reducing the secondary accretion radius. We expect a smaller accretion radius of the secondary black hole to reduce $\dot{M}_{\rm s}$ and increase the fraction of mass expelled.
We set the accretion radius $R_{\rm acc,s}=0.2$ in most of the simulations. 
According to \citet{2009MNRAS.397..657A}, we could estimate the size of the secondary's accretion radius as its centrifugal radius and find $R_{\rm acc,s}\approx R_{\rm H}/3$, where $R_{\rm H}$ is the Hill radius of the black hole. The Hill radius is given approximately by $a(q/3)^{1/3}$. $R_{\rm acc,s}$ is then directly proportional to the separation $a$, so it decreases as the black holes come closer. For our choice of initial conditions, at $a_0=4.75GM_{\rm p}/c^2$ the Hill radius should be $R_{\rm H}\approx 0.33GM_{\rm p}/c^2$. Assuming that the secondary accretion radius is of order the centrifugal radius, we lower $R_{\rm acc,s}$ from 0.2 to 0.05 and to 0.01 in order to resolve it during the secondary's inspiral. 
Considering \emph{ref} with $R_{\rm acc,s}=0.2$ as starting point and then we decrease the secondary's accretion radius to $R_{\rm acc,s}=0.05$ in \emph{sec\_1} and to $R_{\rm acc,s}=0.01$ in \emph{sec\_2}. Details are given in Table \ref{tab:simoverview}. All three simulations have a resolution of $5\cdot 10^5$ particles. 
We point out that it is not possible to lower the accretion radius too much because it is one of the causes for the slow-down of the code. By lowering $R_{\rm acc,s}$ there would be many particles in the vicinity of the secondary black hole that evolve on very short dynamical times. The code will consequently become slower because it has to evolve on smaller time steps to follow the evolution of such particles.
\begin{figure}
	\centering
        \includegraphics[width=\columnwidth]{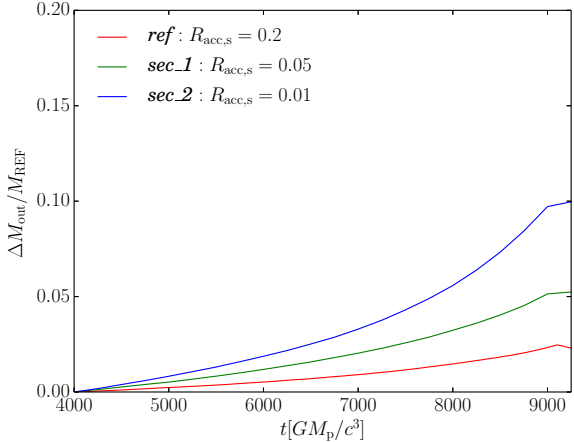}
	\caption[Outer disc mass evolution different secondary accretion radius. ]{Outer disc mass evolution in the last part of the simulation. The fraction of mass funnelled to the outer disc increases for decreasing $R_{\rm acc,s}$. $\Delta M_{\rm out}$ is expressed in unit of $M_{\rm REF}$, the inner disc mass at time $t=4000$. }
	\label{fig:compout_racc2}
\end{figure}
\begin{table}
 \centering
 \begin{tabular}{l | c | c | c }
 \hline
 SIMULATION & \emph{ref} & \emph{sec\_1} &\emph{sec\_2}  \\%[4mm]
 \hline
 $R_{\rm acc,s}$ & 0.2 & 0.05 & 0.01 \\ [1mm]
 $M_{\rm in}/M_{\rm REF}$ & $7.88 \cdot 10^{-5}$ & $6.34 \cdot 10^{-5}$ & $8.42 \cdot 10^{-5}$ \\[1mm]
 $\Delta M_{\rm out}/M_{\rm REF}$ & $2.30 \cdot 10^{-2}$ & $5.24 \cdot 10^{-2}$ & $9.97 \cdot 10^{-2}$ \\[1mm]
 $\Delta M_{\rm acc,p}/M_{\rm REF}$ & $0.698$ & $0.704$ & $0.710$ \\[1mm]
 $\Delta M_{\rm acc,s}/M_{\rm REF}$ & $0.279$ & $0.243$ & $0.190$ \\[1mm]
 \hline
 \end{tabular}
 \caption[Disc masses for different values of secondary accretion radius]{Disc masses at time $t=9250$, for different values of $R_{\rm acc,s}$. $M_{\rm REF}$ is the inner disc mass at time $t=4000$ for every simulation. $M_{\rm in}$ is the inner disc mass. $\Delta M_{\rm out}$ is the increase in the outer disc mass starting from time $t=4000$. $\Delta M_{\rm acc,p}$ and $\Delta M_{\rm acc,s}$ are the mass accreted by the primary and by the secondary in the time interval considered.}
	\label{tab:racc2delta}
\end{table}

Table \ref{tab:racc2delta} lists the fractions of the mass inside and outside the secondary's orbit and the accreted mass by single black holes at time $t=9250$. Masses are expressed in unit of the inner disc mass $M_{\rm REF}$ at time $t=4000$. The fraction of inner disc mass is very low ($M_{\rm in}/M_{\rm REF}=7.88 \cdot 10^{-5}$ in \emph{ref}) because at time $t=9250$ the binary is close to the final merger. Most of the disc mass $M_{\rm REF}$ has been accreted by the black holes and only a small fraction has been channelled outside (third row).
Not surprisingly, the mass funnelled to the outer disc $\Delta M_{\rm out}$ increases by decreasing $R_{\rm acc,s}$, as seen in Figure \ref{fig:compout_racc2}, where $\Delta M_{\rm out}$ is expressed in unit of $M_{\rm REF}$.
$\Delta M_{\rm out}/M_{\rm REF}=2.30 \cdot 10^{-2}$ for \emph{ref} and grows to $9.97 \cdot 10^{-2}$ for \emph{sec\_2}.

Reducing $R_{\rm acc,s}$ has little effect on $\dot{M}_{\rm p}$ ($\Delta M_{\rm acc,p}/M_{\rm REF}$ does not change notably) and if anything increases the primary rate. A larger outer disc mass, by constrast, decreases the secondary accretion rate, since the additional expelled mass is the mass that is not accreted by the smaller black hole.

%%%%%%%%%%%%%%%%%%%%%%%%%%%%%%%%%%%%%%%%%%%%%%%%%%%%%%%%%%%%%%%%%%%%%%%%%%%%%%%%%%%%%%%%%%%%%%%%%%%
%%%%%%%%%%%%%%%%%%%%%%%%%%%%%%%%%%%%%%%%%%%%%%%%%%%%%%%%%%%%%%%%%%%%%%%%%%%%%%%%%%%%%%%%%%%%%%%%%%%
\section{Discussion} \label{sec:commanddisc}

%%%%%%%%%%%%%%%%%%%%%%%%%%%%%%%%%%%%%%%%%%%%%%%%%%%%%%%%%%%%%%%%%%%%%%%%%%%%%%%%%%%%%%%%%%%%%%%%%%%
%%%%%%%%%%%%%%%%%%%%%%%%%%%%%%%%%%%%%%%%%%%%%%%%%%%%%%%%%%%%%%%%%%%%%%%%%%%%%%%%%%%%%%%%%%%%%%%%%%%
\subsection{Comparison with previous work}
\citet{2002ApJ...567L...9A} show the evolution of a supermassive black hole binary with $M_{\rm p}=5\cdot 10^8 M_\odot$ and $q=0.02$, considering a combination of disc driven migration and gravitational decay in a one dimensional code. In this idealized situation, the final stages of the merger drive rapid accretion of the inner disc, reaching an enormous accretion rate of $10^6 M_\odot /$yr. 

The work was revisited by \citet{2012MNRAS.423L..65B}, who improved the set up by using a two dimensional hydrodynamical code. They considered a binary with the same mass ratio of \citet{2002ApJ...567L...9A} to facilitate comparison. 
\citet{2012MNRAS.423L..65B} show that the increase in the outer disc mass $\Delta M_{\rm out}$ is about $80\%$ of the mass in the inner disc and that only $20\%$ is accreted by the primary. This small quantity does not cause any increase in the accretion luminosity and it does not excite the formation of spikes in the surface density of the disc. 

By contrast, all our thin disc simulations confirm that if gas is present between the black holes it is squeezed and rapidly accreted by the primary during the decay. 
We attribute the difference with respect to \citet{2012MNRAS.423L..65B} mostly to an effect related to the disc thickness. \citet{2012MNRAS.423L..65B} adopt $H/R=0.08$ (evaluated at the initial separation $a_0$). Our reference case has $H/R=0.008$ (evaluated at $a_0$), an order of magnitude smaller. 
Indeed we find that, as $H/R$ is increased, more mass tends to flow past the secondary orbit into the outer disc. This is expected, since for larger $H$ pressure forces are stronger and tend to overcome the gravitational torque of the secondary. 
At a given mass ratio, for smaller aspect ratio, the wakes excited by the secondary are more tightly wound; they therefore deposit their energy and angular momentum in a region of the disc (``shocked region'') that is closer to the secondary and that has smaller radial extent \citep{2002ApJ...572..566R}. Upon approaching the secondary, the gas in the inner disc close to the horseshoe separatrix then undergoes a larger deflection inwards which limits the amount of gas that can undergo outward U-turns relative to the secondary.
We cannot directly compare our results to \citet{2012MNRAS.423L..65B} because for the large $H/R$ used by them the whole disc would be accreted before the merger (see our simulations \emph{asp\_3} and \emph{asp\_2}).
However, private communication with the author confirmed us that 2D hydrodynamical simulations with the same code as in \citet{2012MNRAS.423L..65B} and the initial conditions of our \emph{ref} run (thin disc and small mass ratio) give very similar results. This strengthens the hypothesis that the squeezing phenomenon mostly depends on the aspect ratio rather than 3D vs 2D or the secondary accretion radius.

In contrast to \citet{2012MNRAS.423L..65B} and \citet{2002ApJ...567L...9A}, we allow accretion onto the secondary, meaning that we consider particles falling inside the secondary accretion radius to be accreted, even if we do not add their mass and momentum to the black hole. This means that some of the mass that would flow through the secondary's orbit will be captured by the black hole.
Exploring the effects of changing the accretion properties of the secondary, we find (Section \ref{sec:varysecaccrrad}) that it does not affect how much mass flows to the outer disc, but only whether it is accreted or not by the secondary. The primary accretion rate remains essentially unchanged.

Our choice of the disc aspect ratio in the inner disc was dictated by the previous work of \citet{2009MNRAS.398.1392L} and \citet{2015MNRAS.449.1118T}. Figure 4 of \citet{2009MNRAS.398.1392L} clearly show that in the circumprimary disc it is of the order of $10^{-3}$, while in the outer disc it is an order of magnitude larger. We tried to extrapolate a trend for $H/R$ from the work of \citet{2015MNRAS.449.1118T} and we found that the disc aspect ratio is nearly constant around $10^{-4}-10^{-3}$ in the parameter space explored in the paper.
Thus we expect merging binaries to have thinner discs, as in our case, and to give a strong electromagnetic signal from the squeezing of the inner disc.

In the model by \citet{2009MNRAS.398.1392L}, no accretion is allowed from the circumbinary disc onto the circumprimary disc, which leads to mass accretion rates through the circumprimary disc of $\sim 0.001\dot{M}_{\rm Edd}$. If the circumprimary disc is replenished significantly above this value, this would result in a larger value of $H/R$.

%%%%%%%%%%%%%%%%%%%%%%%%%%%%%%%%%%%%%%%%%%%%%%%%%%%%%%%%%%%%%%%%%%%%%%%%%%%%%%%%%%%%%%%%%%%%%%%%%%%
%%%%%%%%%%%%%%%%%%%%%%%%%%%%%%%%%%%%%%%%%%%%%%%%%%%%%%%%%%%%%%%%%%%%%%%%%%%%%%%%%%%%%%%%%%%%%%%%%%%
\subsection{Physical units} \label{sec:physicalunits}
\begin{table*}
 \centering
 \begin{tabular}{l | l | c | c | c }
 \hline 
 SIMULATION & time $t$ & $\dot{M}_{\rm p}/M_{\rm disc}$ & $\dot{M}_{\rm p}/ (M_\odot / {\rm yr})$ & $\dot{M}_{\rm p}/\dot{M}_{\rm Edd}$ \\
 \hline 
% \emph{ref} & $4000$ & $3.64 \cdot 10^{-5}$ & $2.33$ & $0.52$ \\
%  & $8572$ (first spike) & $4.90 \cdot 10^{-4}$ & $31.40$ & $6.98$ \\
%  & $8950$ (second spike) & $4.88 \cdot 10^{-4}$ & $31.27$ & $6.95$ \\
% \hline
%  \emph{resol\_1} & $4000$ & $3.11 \cdot 10^{-5}$ & $1.99$ & $0.44$ \\
%  & $8612$ (first spike) & $7,48 \cdot 10^{-4}$ & $47.93$ & $10.65$ \\
%  & $8983$ (second spike) & $1.04 \cdot 10^{-3}$ & $66.64$ & $14.81$ \\
% \hline
  \emph{resol\_2} & $4000$ & $2.9 \cdot 10^{-5}$ & $1.86$ & $0.41$ \\
  & $8608$ (first spike) & $1.1 \cdot 10^{-3}$ & $70.5$ & $15.67$ \\
  & $9007$ (second spike) & $2.14 \cdot 10^{-3}$ & $137.39$ & $30.53$ \\
 \hline
 \end{tabular}
 \caption[Primary accretion rates in physical units]{Primary accretion rates at the reference time $t=4000$ and at the times of the two spikes for simulation \emph{resol\_2}. The third column shows the dimensionless $\dot{M}_{\rm p}/M_{\rm disc}$, the fourth column shows the accretion rate in units of $M_\odot / {\rm yr}$, the fifth column shows $\dot{M}_{\rm p}$ in units of the Eddington rate $\dot{M}_{\rm Edd}=4.5M_\odot /\rm yr$. The accretion rates at the spikes are larger than Eddington. }
	\label{tab:physicalunits}
\end{table*}
In our discussion so far we have used dimensionless code units or normalised quantities; here we switch to physical units to interpret the results in a physical scenario.
We compare our results for the accretion rates to the critical Eddington value $\dot{M}_{\rm Edd}$. We assume that the luminosity $L$, generated by the accretion of gas by the binary, is given by the formula
\begin{equation}
L=\eta \dot{M} c^2,
\end{equation}
where $\eta$ is the efficiency coefficient. The critical Eddington luminosity and associated Eddington rate are given by:
\begin{align}
L_{\rm Edd}&=1.3\cdot 10^{38} \left( \frac{M_{\rm p}}{M_\odot} \right) \frac{\rm erg}{\rm s}, \\
\dot{M}_{\rm Edd}&=4.5\cdot 10^{-8} \left( \frac{M_{\rm p}}{M_\odot} \right) \left( \frac{0.1}{\eta} \right) \frac{M_\odot}{\rm yr}.
\end{align}
If we choose $M_{\rm p}=10^8M_\odot$ and $\eta=0.1$ we have $L_{\rm Edd}=1.3 \cdot 10^{46} \rm erg/s$ and $\dot{M}_{\rm Edd}=4.5 M_\odot /\rm yr$.
In our simulations accretion rates are expressed in units of the initial disc mass. To switch to physical units, we multiply these dimensionless quantities by two factors: $M_{\rm disc}/M_{\rm p}$ and $c^3/G$, where the first one represents the chosen value of disc mass in units of the primary mass and the second one is the ratio between the unit of mass and time introduced in Section \ref{sec:overview}.

We choose $M_{\rm p}=10^8M_\odot$ and $M_{\rm s}=qM_{\rm p}$ with $q=10^{-3}$.
\citet{2015MNRAS.449.1118T} estimated, using simple 1D diffusion models, that for a $10^8M_\odot$ primary (but different $q$) the inner disc mass at decoupling is of order $1M_\odot$ (this estimate is much larger than previous estimates of the same quantity by \citet{2010MNRAS.407.2007C}, see \citet{2015MNRAS.449.1118T} for a discussion of the origin of the discrepancy). We thus choose $1M_\odot$ as the initial mass of the inner disc.

Table \ref{tab:physicalunits} shows the primary accretion rates at the reference time $t=4000$ and at the times of the two spikes for simulation \emph{resol\_2}. The accretion rates relative to spikes are higher than the Eddington value; the highest being $\dot{M}_{\rm p}/M_{\rm Edd}=30.53$. 

\citet{2015MNRAS.449.1118T}, neglecting mass leakage from the inner disc, found peak luminosities of order $1-20$ times the Eddington luminosities. 
In our 3D simulations we find comparable results, even with different masses and parameters. This confirms the hypothesis of a strong luminosity outburst before the merger.

For a merger of supermassive black holes in the Virgo cluster, located at a distance $\approx 17$ Mpc from Earth, the absolute luminosity associated to the main spike before the coalescence would be $4.8 \cdot 10^{47}$ erg$/$s, assuming a solar mass disc, $q=10^{-3}$ and a primary mass $M_{\rm p}=10^8M_\odot$. Its apparent bolometric magnitude would be $\approx 0.2$.
This highly energetic and transient state has an evolution time-scale of the order of few days (see Fig. \ref{fig:dmdt12resol2ref}).
The event should be similar in luminosity to tidal disruption events and supernovae; a focused study would give some elements to distinguish its peculiar features.
We expect the emitted signal to cover a wide range of frequencies in the electromagnetic spectrum, in particular in the optical and X-rays. 
\citet{2006MNRAS.372..869D} summarized the possible electromagnetic signature of the coalescence for mass ratios $q> 0.01$ that can be detectable in future X-ray missions. Here we show that, if there is some residual gas, a bright precursor can be also observed for an extreme mass ratio.

%%%%%%%%%%%%%%%%%%%%%%%%%%%%%%%%%%%%%%%%%%%%%%%%%%%%%%%%%%%%%%%%%%%%%%%%%%%%%%%%%%%%%%%%%%%%%%%%%%%
%%%%%%%%%%%%%%%%%%%%%%%%%%%%%%%%%%%%%%%%%%%%%%%%%%%%%%%%%%%%%%%%%%%%%%%%%%%%%%%%%%%%%%%%%%%%%%%%%%%
\section{Summary and conclusions} \label{sec:summandconcl}

\begin{figure}
	\centering
	\includegraphics[width=\columnwidth]{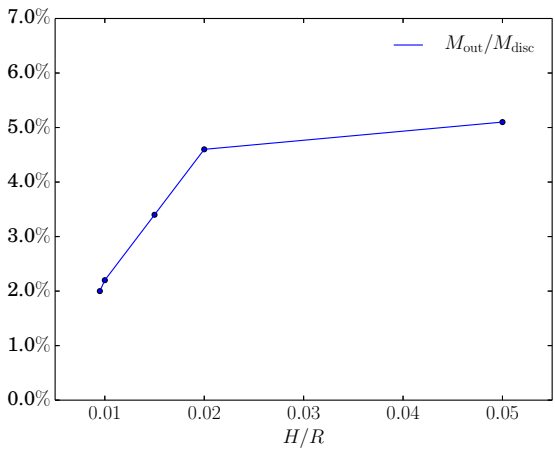}
	\caption[Expelled mass vs disc thickness.]{ Percentage of the expelled mass $M_{\rm out}$ over the initial disc mass $M_{\rm disc}$ versus disc thickness. $H/R$ is computed at $R=R_{\rm in}$. Points represent the five simulations listed in Table \ref{tab:hrcomp_maccmout}.}
	\label{fig:mout_hr}
\end{figure}

We have studied the evolution of gas accretion during the gravitational decay of a supermassive black hole binary, with a mass ratio of $q=10^{-3}$, in order to investigate the possibility of a strong enhancement of luminosity due to the rapid accretion of fossil gas between the black holes in the last phase of coalescence.

For an unequal mass binary, the secondary mass sweeps up gas from its orbit, creating a gap in the disc and acting as a dam. Mass transfer through the gap is not completely suppressed \citep{2013MNRAS.436.2997D,2014ApJ...792L..10D}, since some gas can cross the low density region through horseshoe orbits, as seen in the reference frame of the secondary.

We focused our attention on the case of a supermassive black hole binary with mass ratio $q=10^{-3}$, choosing initial separations $a_0$ smaller than the decoupling radius. 

We found that the secondary black hole rapidly inspirals towards the primary sweeping up the gas in the inner disc. This determines the formation of a narrow spike in the disc surface density that is pushed inwards, while only a small amount of mass is funnelled outside. The inner disc mass is quickly accreted just before the merger, causing a sudden increase in the accretion rate that always exceeds Eddington limit.  

As noticed by \citet{2012MNRAS.423L..65B}, we found that gas particles are channelled to the outer disc through horseshoe orbits as seen in the reference frame corotating with the secondary. 
However, we found that the expelled mass is a negligible fraction of the total initial disc mass. This means that the efficiency of the funnelling mechanism depends on the disc thickness.
By decreasing the disc aspect ratio $H/R$, the total accreted mass with respect to the initial total disc mass increases, while the mechanism of mass funnelling is suppressed. By lowering $H/R$ at $R=R_{\rm in}$ from $0.05$ to $0.01$, the mass funnelled outside ($M_{\rm out}$) decreases from $5\%$ of the total initial disc mass to $2\%$ (see Fig. \ref{fig:mout_hr}). The expulsion tends to flatten for larger $H/R$ because inner disc mass is swallowed before the squeezing takes place.
A larger aspect ratio, assuming the validity of the alpha viscous model, implies a larger kinematic viscosity that causes a faster viscous drift of the inner disc. In our model, the alpha viscosity is fixed by initial conditions through the artificial viscosity mechanism implemented in \textsc{phantom}, but the larger aspect ratio broadens the radial extent of the shocked region and SPH particles are more easily accreted by the central black hole.
On the contrary, when $H/R$ is smaller, the shocked region is thinner and it is located closer to the secondary black hole because wakes are more tightly wound \citep{2002ApJ...572..566R}. Therefore particles close to the inner separatrix of the horseshoe region are more easily deflected inwards rather than outwards.

The dependence on $H/R$ of the squeezing phenomenon is the main reason for the discrepancy with \citet{2012MNRAS.423L..65B}, who considered a thicker disc.
The dependence of the enhancement on disc thickness may provide a method for observationally inferring $H/R$ from measurements of the accretion rate.

The fraction of mass funnelled outside depends also, but more weakly, on the secondary accretion radius.
We find that the primary accretion rate remains nearly unchanged when decreasing the secondary accretion radius, while the secondary accretion rate is reduced. Consequently there is more mass available to be dynamically expelled and this is exactly what we found in our results. However, the increase in the expelled mass is always restricted to a few percent.

The accuracy of our results depends on resolution. Due to the limitation of computational resources, we performed most of our simulations by employing $5\cdot 10^5$ SPH particles. In two cases we raised the resolution to $10^6$ and $2 \cdot 10^6$ particles. We find that low resolution simulations tend to under-resolve the accretion rate, hence our simulations are a lower limit on the true enhancement.
In the high resolution case we find that in the last stages of the merger the accretion luminosity of the binary is increased by a factor of about $70$ with respect to the initial constant one. This corresponds to a super-Eddington luminosity. The primary accretion rate reaches the value of $\dot{M}_{\rm p}=30.53 M_{\rm Edd}$, for a primary mass of $M_{\rm p}=10^8 M_\odot$ and an accretion efficiency $\eta=0.1$.

We limited our simulations to a restricted region of parameter space, in particular we considered only the case of mass ratio $q=10^{-3}$. A wider investigation would require more computational resources. 

It would be useful to compare our results with general relativistic simulations that account for the distorted geometry near the binary. We could then understand if it is safe to neglect relativistic correction to the potential of the binary as we have done in our work. The possibility of such a comparison already exists \citep{2014PhRvD..90j4030G}.

In conclusion, with our three dimensional simulations we predict that electromagnetic signal from a coalescing binary should be detectable.
Even if the study was carried out with low resolution and we investigated only the case of $q=10^{-3}$, the results are interesting because we find that only a small amount of inner disc mass can be funnelled outside, contrary to previous results.
Of course the study could be improved by increasing the resolution and by investigating the parameter space widely. This will allow to understand the phenomenon in the case of various combinations of mass ratio and disc thickness, in view of a possible detection in future surveys.

%%%%%%%%%%%%%%%%%%%%%%%%%%%%%%%%%%%%%%%%%%%%%%%%%%%%%%%%%%%%%%%%%%%%%%%%%%%%%%%%%%%%%%%%%%%%%%%%%%%
%%%%%%%%%%%%%%%%%%%%%%%%%%%%%%%%%%%%%%%%%%%%%%%%%%%%%%%%%%%%%%%%%%%%%%%%%%%%%%%%%%%%%%%%%%%%%%%%%%%
\section*{Acknowledgments}

We thank James Wetter and Matt Stuart for their initial work on the gravitational wave decay in \textsc{phantom} and Chris Nixon for useful discussions.
We acknowledge the referee, Cl\'ement Baruteau, for his useful comments during the revision of the manuscript.
Figures \ref{sim7_snapshotmovie} and \ref{fig:sim7_horseshoe} were produced using SPLASH \citep{2007PASA...24..159P}, a visualization tool for SPH data.

%%%%%%%%%%%%%%%%%%%%%%%%%%%%%%%%%%%%%%%%%%%%%%%%%%



\bibliographystyle{mnras}
\bibliography{biblio} 

%%%%%%%%%%%%%%%%%%%%%%%%%%%%%%%%%%%%%%%%%%%%%%%%%%
%%%%%%%%%%%%%%%%%%%%%%%%%%%%%%%%%%%%%%%%%%%%%%%%%%

% Don't change these lines
\bsp	% typesetting comment
\label{lastpage}
\end{document}